\documentclass{elsart}
\usepackage{graphicx}

\begin{document}

\begin{frontmatter}
\title{Characterization of agostic interactions in theory and computation}
\author{Matthias Lein}
\address{Centre for Theoretical Chemistry and Physics, New Zealand Institute for Advanced Study, Massey University Auckland}

\ead{m.lein@massey.ac.nz}

\tableofcontents

\begin{abstract}
Agostic interactions are covalent intramolecular interactions between an electron deficient metal and a $\sigma$-bond in close geometrical proximity to the metal atom. While the classic cases involve CH $\sigma$-bonds close to early transition metals like titanium, many more agostic systems have been proposed which contain CH, SiH, BH, CC and SiC $\sigma$-bonds coordinated to a wide range of metal atoms.
Recent computational studies of a multitude of agostic interactions are reviewed in this contribution. It is highlighted how several difficulties with the theoretical description of the phenomenon arise because of the relative weakness of this interaction. The methodology used to compute and interpret agostic interactions is presented and different approaches such as \emph{atoms in molecules} (AIM), \emph{natural bonding orbitals} (NBO) or the \emph{electron localization function} (ELF) are compared and put into context. A brief overview of the history and terminology of agostic interactions is given in the introduction and fundamental differences between $\alpha$, $\beta$ and other agostic interactions are explained.
\end{abstract}


\end{frontmatter}

\section{Introduction}

\begin{figure}[h] 
   \centering
   \includegraphics[scale=.25]{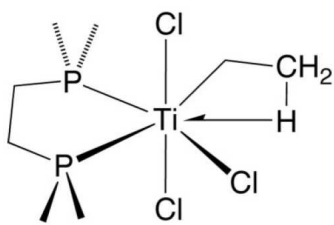} 
   \caption{The first "taylor made" compound exhibiting an agostic interaction. (Taken from ref. \cite{Green-2007-PNAS}).}
   \label{fig:first}
\end{figure}

Transition metal compounds which exhibit a close proximity of CH systems to the metal atom were discovered relatively early in the mid 1960's and early 1970's as the quality and availability of x-ray crystallography improved \cite{PlacaIbers-1965-InorgChem,Trofimenko-1968-JACS,Trofimenko-1970-InorgChem,Cotton-1972-InorgChimActa,Cotton-1974-JACS,Cotton-1974-JChemSocChemComm}. But the dispute as to whether or not the close approach was caused by an attractive force between the $\sigma$-bond and the metal atom could not be resolved at that time. It was not until the 1980's that Green and coworkers tried to systematically conceive and prepare a compound that would unambiguously identify such an attractive force \cite{Green-1986-Dalton,Green-1982-JChemSocChemComm}. This compound (shown in Fig. \ref{fig:first}) was an attractive target for synthesis because of several reasons.
Firstly, the compound features an electron deficient early transition metals with empty d-orbitals which could accept charge-donation from a nearby donor such as the conveniently close methyl group. Secondly, the compound is relatively unrestricted in terms of steric crowding.

This made, for the first time, a convincing argument in favor of the existence of a new previously unknown type of interaction. Soon after, Brookhart and Green published their two early landmark reviews \cite{Brookhart-Green-JOrganometChem-1983,Brookhart-Green-1988-ProgInorgChem} on the subject which are today the cornerstones and starting points for any investigation into agostic interactions.

The very term \emph{agostic} (from greek $\alpha\gamma$o$\sigma\tau$o$\varsigma$, meaning "to draw towards") was coined by the same pioneering authors \cite{Brookhart-Green-JOrganometChem-1983} and was originally only meant to describe 3-center-2-electron bonds involving CH groups while the term \emph{non-agostic} is used for systems which exhibit a similar close proximity of a metal atom and a $\sigma$-bond but are better described in terms of hydrogens bonding or similar mechanisms or system that could be expected to exhibit an agostic shortening of bond lengths but do not show such behavior. The terms \emph{anagostic} \cite{Lippart-1990-JACS} or \emph{pseudo-agostic} \cite{Desiraju-1997-Organometallics} are also sometimes used.

Ever since the term \emph{agostic} was coined there has been a steady increase in the number of compounds which display agostic interactions as well as a steady rise in papers describing these compounds \cite{Green-2007-PNAS}. While at first considered to be rare and somewhat special, we now know that agostic interactions are, in fact, rather common and provide very important stabilization of transition metal compounds \cite{Perutz-Etienne-2007-AngewChem}.

Many approaches can be used to determine if an agostic bond is, in fact, present. Traditionally, geometrical parameters have always been the first indication that an agostic interaction was present. In order to distinguish between proper agostic interactions and other weak interactions such as hydrogen bonds several distance ranges have been established. The metal to hydrogen distance is usually in the range of 1.8 -- 2.3 \AA, while the metal-hydrogen-carbon angle is in the range of 90 -- 140 degrees \cite{Green-2007-PNAS}. While this usually has remarkable predictive power, often theoretical methods which analyze the electron density or the wavefunction are needed to decide on difficult cases. Many methods have been used for this purpose and a set of standard procedures is beginning to emerge \cite{Clot-Eisenstein-StructBond-2004,Scherer-McGrady-2004-AngewChem}. This is, of course, complicated by the fact that most intramolecular effects that are known in chemistry are hard to define in an unambiguous way and sometimes several different approaches that have been devised to probe the same system can come up with very different outcomes. Another good example for this kind of dilemma is the source of the rotational barrier in ethane and the question why the staggered rotamer is lower in energy than the eclipsed one \cite{Bickelhaupt-2003-AngewChem,Weinhold-2003-AngewChem}.

\subsection{Agostic Interactions by type}

\begin{figure}[h] 
   \centering
   \includegraphics[scale=.6]{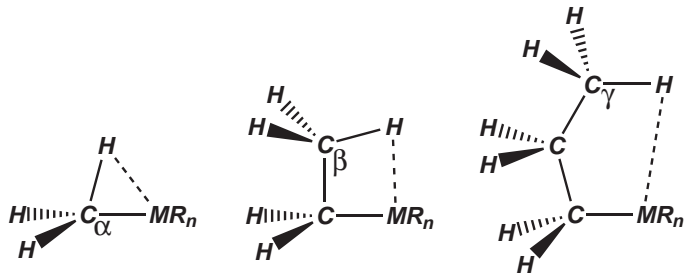} 
   \caption{$\alpha$, $\beta$ and $\gamma$ agostic interactions.}
   \label{fig:a-b-g}
\end{figure}

As mentioned earlier, there are several types of agostic interactions. They are usually grouped according to the connectivity between the interacting atom (usually hydrogen) and the metal along covalent bonds (See fig. \ref{fig:a-b-g}).
In this sense agostic interactions are not categorized by the nature of the actual metal-hydrogen interaction but by the structure of the ligand which chelates the metal. Nevertheless, the simplicity of this approach means that this is now the most common categorization used in the literature. Many systematic studies have been carried out on subsets of $\alpha$, $\beta$ and other agostic interactions and the following sections aim to give an overview of some recent developments in the context of earlier work in this field.

\subsubsection{$\alpha$-agostic interaction}
The first theoretical work on $\alpha$-agostic interactions was done more than 20 years ago by Goddard, Hoffmann and Jemmis \cite{Hoffmann-1980-JACS}. The importance of this work stems from the fact that this study radically altered the view on $\alpha$-agostic interactions. Until then, it was thought that the direct interaction of the occupied bonding orbitals of the carbon-hydrogen bond with vacant orbitals in an electron deficient metal atom facilitate the formation of agostic interactions. Hoffmann and coworkers could show that this is indeed not the case.

\begin{figure}[h] 
   \centering
   \includegraphics[scale=0.8]{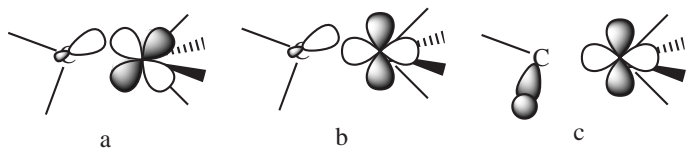} 
   \caption{Orbital interactions in the $\alpha$-agostic case by significance: a) Stabilizing metal-carbon interaction b) de-stabilizing metal-carbon interaction c) stabilizing CH-metal interaction.}
   \label{fig:alpha}
\end{figure}

The model system employed in \cite{Hoffmann-1980-JACS} clearly demonstrated that by tilting the $\rm CH_2$ towards the metal two new interactions are switched on. Firstly, as previously thought, the $\sigma_{\rm CH}$ bonding orbital starts to interact with the metal's LUMO+1 orbital. However, because the energy difference between the two interacting orbitals is large, the $\sigma_{\rm CH}$ bonding orbital is very low in energy, the stabilization through this interaction is very small. Another interaction that is favored by the new, tilted structure is the interaction of the metal's LUMO+1 and the HOMO of the $\rm CH_2$ ligand in its singlet state, an  $\rm sp^2$ lone pair orbital of $\sigma$ symmetry. Because these orbitals are significantly closer to each other in energy and overlap significantly in the tilted geometry, they deliver the main energetic contribution to the agostic interaction. Because the metal's LUMO is not as favorably hybridized as the metal's LUMO+1 the loss of overlap and hence stabilization, by tilting the $\rm CH_2$ ligand's $\sigma$ HOMO away in the primary bonding metal-carbon interaction is more than compensated by this new stabilizing interaction (See fig. \ref{fig:alpha}).

So, even though the previously proposed direct interaction between the metal atom and the carbon-hydrogen bond is present and even accompanied by the usual tell-tale sign of a decreased NMR $\rm ^1J_{C-H}$ coupling constant, it is too small to cause the geometric change by itself. It should rather be seen as a consequence of a change in the metal-carbon interaction.

In a subsequent study it has been shown that this concept can be extended to explain subtle differences in binding of otherwise similar compounds of which some exhibit agostic structures and others don't \cite{Eisenstein-1985-JACS}. Here again, it is the bonding between the empty metal atoms and the HOMO of the ligand which determines if a tilting of the ligand is energetically favorable or not. In the reported titanium compounds an additional dmpe ligand forces the titanium into octahedral coordination and hence distorts the structure of the complex enough for two empty metal orbitals to switch places enabling an effective interaction with the HOMO of the $\rm CH_3$ ligand. The same compound without the additional dmpe ligand has a tetrahedral mode of coordination where the $\rm CH_3$ ligand gains nothing by tilting towards the metal. Note here that the agostic geometry occurs in the sterically more crowded complex while it is absent in the sterically less crowded complex which highlights the electronic origin of agostic interaction as opposed to steric influences. A result that has also been obtained somewhat more thoroughly from the comparison of QM/MM and DFT calculations \cite{Eisenstein-1998-JACS,Eisenstein-1999-JACS}

Such studies are carried out with essentially the same results but different analytical tools and more sophisticated computational models to this date. Some authors go as far as arguing for a replacement of the term \emph{agostic interaction} with \emph{agostic geometry} to emphasize the fact that the structural change in those compounds does not stem from a direct interaction of the metal atom and the carbon-hydrogen bond \cite{Dobado-2006-Organometallics,Dobado-2008-JPhysChemA}.

\subsubsection{$\beta$-agostic interaction}

Not surprisingly many of the conclusions drawn for $\alpha$-agostic interactions are also valid for $\beta$-agostic interactions \cite{Eisenstein-1984-Organometallics,Scherer-1998-JACS,Scherer-1998-Organometallics}. Here too the main contribution to the stabilization comes from the weakening of the $\rm C_\alpha$-metal bond and not from the direct donation of $\rm C_\beta$H-bond density to vacant metal orbitals.
\begin{figure}[h] 
   \centering
   \includegraphics[scale=.55]{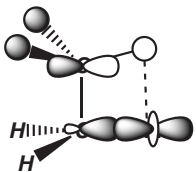} 
   \caption{Orbital interactions in the $\beta$-agostic case. The metal-$\rm C_\alpha$ $\sigma$-bond at the bottom and the $\rm C_\beta$H-metal secondary interaction at the top.  Figure adapted from \cite{Scherer-1998-JACS}}
   \label{fig:beta}
\end{figure} 
The orbital picture is, however, somewhat different (See fig. \ref{fig:beta}). In this case the main driving force for the stabilization of the agostic geometry is the delocalization of the two electrons in the metal-$\rm C_\alpha$-$\rm C_\beta$ system through negative hyperconjugation \cite{Scherer-McGrady-2004-AngewChem}.

\subsubsection{$\gamma$, $\delta$ and \emph{remote}-agostic interactions}

True $\rm C_\gamma$H agostic interactions are rare but not unknown \cite{DenHaan-1986-Organometallics,Yang-2003-JCP}. Although sometimes the potential for the presence of a $\gamma$-agostic interaction is noted, drawn from purely structural arguments based on newly available crystallographic data \cite{PowerBartlett-1988-ChemComm,Collins-Ziegler-2007-Organometallics,Yang-2003-JCP}. It turns out that often a closer inspection of the wavefunction or the electron density does not give evidence to this end \cite{HayPoli-2002-Organometallics,Eisenstein-2003-NewJChem}. In many cases other interactions are energetically more favorable and easily replace the weak $\gamma$-agostic interaction -- it has also been argued that the close $\rm C_\gamma$-metal approach is actually associated with a repulsive interaction \cite{HayPoli-2002-Organometallics}.

This highlights one fundamental problem of agostic interactions. While the initial definition and the early work has been done mainly focusing on structural parameters \cite{Brookhart-Green-JOrganometChem-1983,Brookhart-Green-1988-ProgInorgChem} the underlying physical reasons for the CH approach to a metal atom can be diverse \cite{Clot-Eisenstein-StructBond-2004}. Already, interactions of groups like NH, where the hydrogen atom is profoundly positively charged, which used to be grouped alongside agostic interactions, are not longer considered part of that group. The reason is simple in this particular case. While classical agostic interactions are characterized by electron density being transferred from a CH bond into a vacant orbital of an electron deficient metal, the NH more closely resembles a classic hydrogen bond, where a relatively electron rich metal atom interacts with a proton \cite{Popelier-1998-JOrganometChem}.

If one follows the literature one can also find $\delta$-agostic \cite{Nielson-2004-Dalton,HayPoli-2003-InorgChem,Davies-Macgregor-2006-Organometallics} and even $\epsilon$-agostic interactions \cite{Ephritikhine-Maron-2006-JACS,Montiel-Palma-Sabo-Etienne-2007-ChemComm}.

\section{Agostic Interactions by method}

Since many methods have been applied to the characterization of agostic interactions it is instructive to review the main computational tools that are commonly used. This section is divided into two subcategories.

Firstly, the purely computational tools are  reviewed.  While density based methods like Bader's AIM\footnote{recently: QTAIM} are, in principle, applicable to experimentally obtained electron densities, this approach is not often seen and hence this method is presented in this section.

Secondly, methods which employ molecular spectroscopy will be looked at. While most spectroscopic properties like vibrational spectra, nmr shifts and coupling constants can be calculated as well, these tools are of utmost importance to experimentally working groups and data is probably easier obtainable experimentally than computationally.

\subsection{Computational Approach}

\subsubsection{AIM - Atoms in Molecules}

The \emph{atoms in molecules} method by Bader \cite{Bader-1990-Book} is an elegant way to rationalize chemical information through physical observation. Since many chemical concepts, starting with the term \emph{chemical bond}, evolved when chemistry was still exclusively empirical it is often hard to translate them into physically meaningful concepts \cite{Frenking-2007-JComputChem}. AIM uses the topology of the electron density as a basis which has the added advantage of being independent from model quantities such as molecular orbitals or basis functions and, indeed, computation.

AIM looks at the electron density itself and gradient vector field of the electron density of a given system to draw its conclusions. First, critical points in the electron density are located. Maxima in the density are identified as the location of the nuclei\footnote{With the rare exception of \emph{non-nuclear attractors} which are also local maxima in the electron density \cite{Edgecombe-1993-ZeitschrFNaturForsch}}. Saddle points (points of inflection) are then identified as features of chemical bonding. So called \emph{bond critical points} (points of inflection with respect to one coordinate) are located on gradient paths connecting two atoms in a given molecule. The existence of a bond critical point is sufficient evidence for the existence of a chemical bond in AIM theory and some bond properties can be deduced from properties of the electron density at this point. \emph{Ring critical points} and \emph{cage critical points} are points of inflection with respect to two and three coordinates respectively. Ring critical points often feature in cases of agostic interactions as well because the chelating nature of the ligand often creates an almost planar ring of atoms.

\begin{figure}[h] 
   \centering
   \includegraphics[scale=.3]{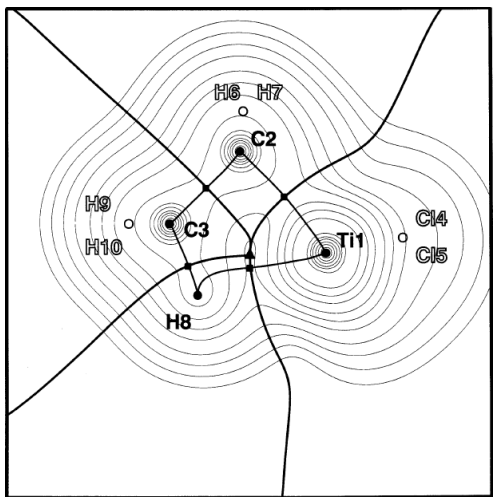} 
   \caption{Graph of the electron density of $\rm Cl_2Ti$-$\rm CH_2CH_3^+$ in the Ti-$\rm C_\alpha$-$\rm C_\beta$-$\rm H_\beta$ plane. The bond critical points are denoted by solid squares and the ring critical point is given by a solid triangle. Taken from \cite{Popelier-1998-JOrganometChem}}
   \label{fig:aim}
\end{figure}

The first systems to undergo an AIM inspection were $\rm Cl_2Ti$-$\rm CH_3^+$, $\rm Cl_2Ti$-$\rm CH_2CH_3^+$ and $\rm Cl_2Ti$-$\rm CH_2CH_2CH_3^+$ with particular attention to the difference between hydrogen bonds and agostic interactions \cite{Popelier-1998-JOrganometChem}. The electron density of $\rm Cl_2Ti$-$\rm CH_2CH_3^+$ is shown in fig \ref{fig:aim}.

The structural parameters of $\rm Cl_2Ti$-$\rm CH_2CH_3^+$ clearly suggest the presence of an agostic interaction. The $\rm C_\beta$-$\rm H_\beta$ distance is with 1.145 \AA~rather short and the Ti-$\rm C_\alpha$-$\rm C_\beta$ angle is with 84.9 degrees unusually acute. More importantly, the graph shows a bond critical point and a bond path connecting the Ti atom and $\rm H_{beta}$, proving the existence of a direct interaction between those two atoms. The question is now whether the close proximity is due to a classical hydrogen bond or a true agostic interaction.
To ascertain the nature of the interaction between the $\rm H_{beta}$ and Ti several other elements of AIM theory can be employed.

The electron density of the Ti-H bond critical point can be used to determine the bond order and hence gives some indication of the bond strength \cite{Bader-1987-JACS}. In the Ti compounds studied the electron density amounts to 0.04--0.05 a.u. which indicates a weak interaction as expected. However, the established values for hydrogen bonds lie in the range of 0.002--0.035 a.u. \cite{Popelier-1995-JPhysChem} indicating that the interaction in these compounds is fundamentally different. Similarly, the Laplacian $\nabla^2\rho$ of the electron density at the bond critical point falls outside the established region for hydrogen bonds. Finally, the AIM charge of the agostic hydrogen atom is integrated to slightly negative values of the order of -0.1 e as opposed to slightly positive values for the same hydrogen atom if the $\rm CH_3$-group is rotated away from the metal atom. This is exactly to opposite behavior than one would expect from a hydrogen bond where the AIM charge of the hydrogen atom changes sign in the opposite direction upon formation of the bond \cite{Bader-1983-JACS}. This criterion has been successfully used to identify a misassigned CH-Cu agostic interaction \cite{Desiraju-2006-ChemComm}. While the study of systems of the form $\rm Cl_2TiR^+$ in \cite{Popelier-1995-JPhysChem} is now used to benchmark other AIM studies of agostic systems a word of caution should be added. The molecules that could be studied by theoretical methods in 1995 do not exist in the laboratory and as such can only be used for a qualitative calibration of the method. The difference between the electron donating capabilities of Cl as opposed to the Cp in the real system could well mean that the qualitative picture changes once the experimentally accessible system is studied. A similar point is made by Scherer, McGrady and coworkers in their first combined experimental and theoretical AIM study of an agostic system \cite{Scherer-McGrady-1998-ChemCommun}. This also marks the first time where an experimentally obtained electron density had been used for an AIM analysis of an agostic compound.

Even though $\rm Cl_2Ti$-$\rm CH_3^+$ exhibits a small Ti-$\rm C_\alpha$-$\rm H_\alpha$ angle of just under 90 degrees, which is typical for $\alpha$-agostic geometries, no bond critical point between Ti and $\rm H_\alpha$ can be found. This is in line with earlier observations indicating that $\alpha$-agostic geometries are mostly driven by changes in the metal-$\rm C_\alpha$ bond rather than an actual CH-metal interaction. However, caution should be taken with systems that exhibit particularly weak agostic interactions as it can be difficult to identify bond paths and bond critical points which could lead to mis-identification \cite{Coppens-2003-JACS}. Only recently the first bond-critical point between an $\alpha$-agostic H atom and a metal center has been observed \cite{Ruiz-2008-Organometallics}. Although the bond path connects the hydrogen atom of the agostic CH group and the \emph{second} metal atom of a $\rm Mo_2$ bi-metallic center. While this is certainly an unusual topological situation it is fair to ask whether this is truly an $\alpha$-agostic arrangement.

Scherer and coworkers compared the experimental and calculated electron density in an AIM study of Et-Ti(Cl)$_3$(dmpe) \cite{Scherer-1998-ChemComm}. Reassuringly it was found that the topology of both densities was so similar that the same number of bond critical points and ring critical points was found. Furthermore more evidence was presented that a significant contribution to the stabilization of agostic structures comes from the change in the metal-$\rm C_\alpha$ bond. And a criterion relating the curvature of this bond to $\beta$-agostic interaction was offered. See also \cite{Scherer-McGrady-2004-AngewChem}.

AIM theory has also been used in the investigation of the propagation barrier of ethylene polymerization with titanium compounds \cite{Duarte-Heine-2006-Theochem} -- the prototypical agostic system. Duarte, Heine and coworkers confirm the Brookhart Green mechanism of polymerization where the formation of a $\pi$ complex is followed by a transition state which is stabilized by an agostic interaction \cite{Brookhart-Green-JOrganometChem-1983}. They state that while there is no bond critical point indicating the existence of agostic interactions in this system the molecular orbital picture leads to the conclusion that the $\beta$-agostic group acts as a two-electron ligand. This means that the first step of the polymerization is dominated by the CH$\cdots$Ti interaction, while the second step is dominated by a $\rm C_\beta\cdots$Ti interaction.

\begin{figure}[h] 
   \centering
   \includegraphics[scale=.4]{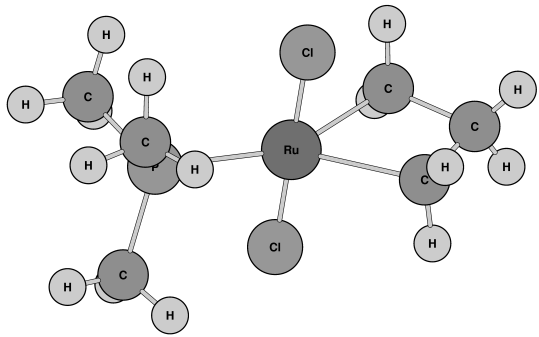} 
   \caption{$\alpha$-$\beta$-CCC agostic interaction. Adapted from \cite{Suresh-2006-JOrganometChem}}
   \label{fig:alpha-beta-CCC}
\end{figure}

Another interesting type of bonding where AIM theory has been employed has been termed $\alpha\beta$-CCC agostic bonding \cite{Suresh-2004-Organometallics,Suresh-2005-Dalton,Suresh-2006-JOrganometChem,Jacobsen-2006-Dalton}. Here, in a metalla cyclobutane complex only the $\beta$-carbon atom is not directly bound to the metal but still exhibits a close metal-C approach combined with significant elongation of the $\rm C_\alpha$-$\rm C_\beta$ single bonds (See fig \ref{fig:alpha-beta-CCC}).
Because of the cyclic arrangement of the cyclobutane ligand no bond critical point between the $\rm C_\beta$ atom and the metal is found. Not surprisingly, the local topology only shows a ring critical point which leads Suresh to the conclusion that the interaction is mainly $\pi$-type without an accompanying $\sigma$ component. An analysis of the electron density at the bond critical points of the $\rm C_\alpha$-$\rm C_\beta$ bonds fully support the agostic type of bond here. The reduced electron density of those bonds combined with the diminished covalent character paint a conclusive picture.
It was also noted that complexes exhibiting $\alpha$-$\beta$-CCC agostic interactions could be considered as the first examples of metal-carbon $\pi$-interactions without concurrent $\sigma$-interaction.

A somewhat unusual PC-Pd agostic interaction was reported recently by McGrady, Vilar and coworkers \cite{McGrady-Vilar-2006-Organometallics}. In this study the electronic structure of a palladium dimer complex is highly dependent on the ligand. The substitution of the tBu groups of the original ligand by hydrogen atoms changes the picture significantly. one bond critical point which was present in the original system disappears and a new one emerges between the metal atom and a remote phosphorous atom. In this new system the interaction between the metal and the phosphine ligand is dominated by a PC-agostic interaction which is rather distinct from the common metal-C=C interaction of the original system.

The AIM picture provides evidence for a direct interaction between the agostic hydrogen atom and the metal at least for $\beta$ and $\gamma$ agostic systems. Evidence for the classical picture where the CH bond donates density into vacant orbitals of an electron deficient metal is not found. In this case there should be, in principle, a bond critical point and consequently a bond path between the metal atom and the C-H bond critical point \cite{Clot-Eisenstein-StructBond-2004} to indicate such an interaction. While technically true, it should be noted that it is topologically impossible for a bond path to connect two bond critical points which are, by definition, points of inflection with respect to only one coordinate.

One of the shortcomings of the currently available density functional methods is that in some cases different functionals not only give different quantitative answers, but sometimes the picture can even change qualitatively. The area of agostic interactions is no exception here. McGrady and coworkers have recently pointed out that only certain density functionals that conform to the uniform electron gas (UEG) limit accurately reproduce the electron density in the area of a $\beta$-agostic bond \cite{McGrady-2008-Organometallics}. Thus, only these functionals will yield the correct agostic geometry on which an AIM analysis can be based. Other functionals may fail to locate the agostic minimum entirely. McGrady and coworkers point out that only a functional that correctly reproduces a range of physical observables is fit for an analysis of agostic bonding situations.

\subsubsection{NBO - Natural Bond Orbitals}

One of the most widely used methods for the investigation of chemical bonding is Weinhold's \emph{natural bonding orbital} approach \cite{WeinholdLandis-2005-Book}. NBO generalizes the concept of natural orbitals, which are obtained by diagonalizing the first-order density matrix, by deriving \emph{natural atomic orbitals} and \emph{natural bond orbitals} from the one-electron density matrix. In this way it is possible to define atomic orbitals and hence atomic charges for populations analyses as well as  chemical bonds which are derived from electron density between atoms. One particular desirable feature of \emph{natural population analysis} is that it is numerically more stable than other comparable analyses such as the Mulliken scheme which can give diverging results for larger basis sets.

Of particular use in coordination chemistry is another feature of NBO theory. Donor-acceptor relationships can be estimated through perturbation theory. By reformulating perturbation theory in terms of orbitals, orbital energies and orbital occupations it is possible to derive a second order term which represents the two electron stabilizing interaction of an occupied orbital with a vacant orbital. This term, $\Delta E_{i\rightarrow j^*}^{(2)}$, is frequently used in the investigation of agostic interactions.

\begin{figure}[h] 
   \centering
   \includegraphics[scale=.6]{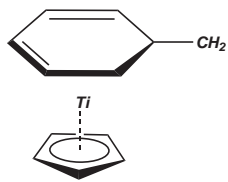} 
   \caption{$\rm Ti(C_5H_5)^+$ and the 5-methylene-1,3-cyclohexadiene anion as calculated in \cite{GleiterErnst-1998-JACS}}
   \label{fig:1st-nbo}
\end{figure}
The first system which was subjected to NBO treatment was a sterically crowded 14 electron Ti compound which exhibited unusually short distances between CC bonds of saturated carbon atoms and the metal atom \cite{GleiterErnst-1998-JACS}. In the model system the interactions between $\rm Ti(C_5H_5)^+$ and the 5-methylene-1,3-cyclohexadiene anion were calculated (See fig \ref{fig:1st-nbo}). The  natural bond orbitals showed significant occupancy decreases for several C-C bonds and one C-H bond consistent with donation of electron density into vacant Ti orbitals. At the same time the second order perturbation energy indicates a large stabilizing contribution (57 kcal/mol) from donor-acceptor interaction between the four C-C bonds and the Ti atom.

Almost at the same time another NBO study by Kaupp on tungsten and rhenium d$^0$ and d$^1$ complexes appeared \cite{Kaupp-1998-ChemEurJ}. These complexes of the $\rm MOR_4$ type (R = H, $\rm CH_3$) are highly fluxional with many equivalent structures rapidly interconverting into each other via low-lying transition states. Surprisingly, internal rotation of the methyl groups is severely restricted by relatively strong agostic interactions of the $\alpha$-hydrogen atoms with the metal atom and distorted structures of low symmetry are favored over the symmetrical regular square pyramidal form. Second order perturbation theory suggests that this is at least partly caused by agostic - in the original also called hyperconjugative - interactions of the $\rm CH_3$ ligand which has to tilt in order to interact with the metal atom.

\begin{figure}[h] 
   \centering
   \includegraphics[scale=.3]{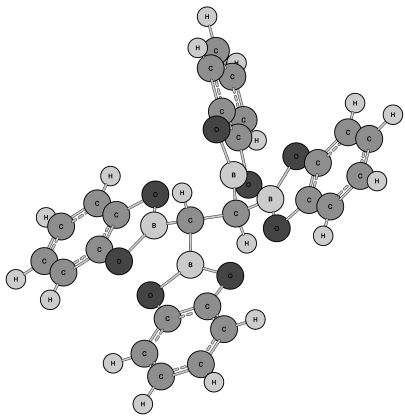} 
   \caption{CH$\cdots$B agostic interaction discussed in \cite{GleiterSiebert-1999-EurJInorgChem}. The catechol ligands tilt to enable a stronger interaction.}
   \label{fig:ch-b-agostic}
\end{figure}
The first study to report an NBO analysis on a metal free system investigated the structure and bonding of borylethanes and borylethenes \cite{GleiterSiebert-1999-EurJInorgChem}. Although not a classical agostic system, the authors employ the same terminology and apply it in their discussion of a twisted tetraborylethane. Here, the second order perturbative  treatment was used to elucidate the interaction between the formally unoccupied boron $\rm p_z$ orbital and the ethyl CH bonds. The results in this case show that there is a significant build-up of density at the boron atom. The $\rm p_z$ orbital has a natural population of 0.318 e and $\Delta E_{i\rightarrow j^*}^{(2)}$ for the CH-$\sigma\rightarrow$B $\rm p_z$ donation yields 8.96 kcal/mol.
A more traditional agostic interaction with a $\gamma$-BH$\cdots$Sn structure was recently reported \cite{Izod-2006-Organometallics}. Here, the B-H $\sigma$ bond orbitals interact strongly with the empty Sn 5$\rm p_x$ orbital of the main group element. The second order perturbation treatment assigns 20 to 30 kcal/mol to this interaction depending on the diastereomer.
 
Lanthanide and actinide complexes have also been reported \cite{HayPoli-2003-InorgChem,Ephritikhine-Maron-2006-JACS,Niemeyer-2006-InorgChem,Li-Andrews-2007-AngewChem,Andrews-Li-2008-InorgChem}. Brady et al probed the interactions in the samarium compound $\rm Sm[N(SiMe_3)_2]_3$ \cite{HayPoli-2003-InorgChem} where one of the methyl groups of each amido ligand exhibits a close approach to the metal atom, possibly a $\gamma$-agostic interaction. The authors chose a combined molecular orbital and NBO approach for their investigation of this problem. By deleting the valence d-functions from their basis-set they tried to elucidate the degree of covalent and electrostatic interactions in their system. Consistent with the assumption that agostic interaction are mainly covalent interactions they found significant structural changes after the removal of the valence d-functions consistent with the presence of a $\beta$-SiC agostic interaction. Population analyses confirmed that the interacting atoms had slightly higher negative charges as those not involved in the agostic bond. Interestingly, evidence for the presence of a $\gamma$-agostic interaction could not be established even though the purely structural appearance suggested otherwise.

An interesting $\epsilon$-agostic interaction was reported in a uranum thiolate complex \cite{Ephritikhine-Maron-2006-JACS}.
\begin{figure}[h] 
   \centering
   \includegraphics[scale=.4]{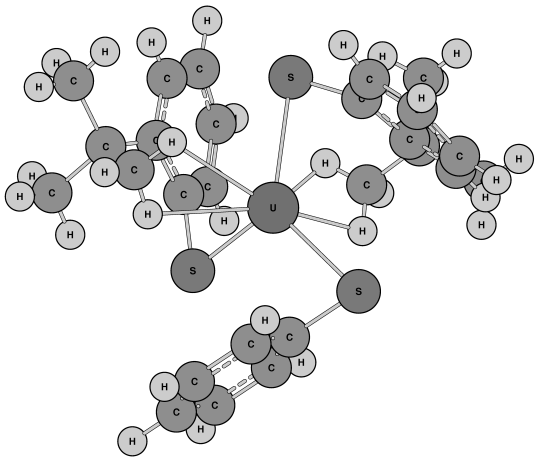} 
   \caption{Interesting $\epsilon$-agostic interaction in a uranium thiloate complex \cite{Ephritikhine-Maron-2006-JACS}}
   \label{fig:eta-agostic}
\end{figure}
Despite the relatively large CH-U distance of almost 2.6 \AA (See fig \ref{fig:eta-agostic}), the authors report a second order stabilization energy $\Delta E_{i\rightarrow j^*}^{(2)}$ for the donation of density of one CH bond of one of the tBu groups into an empty d-f hybrid orbital of uranium of 5 kcal/mol. Given, that the covalent radii of actinides are relatively large an agostic interaction over this distance is possible,but still somewhat unusual. Another look at this system with a density based method such as AIM might yield further insight and could help in providing more evidence for this unusual interaction in this particular system.

A chiral agostic system in the pyramidal actinide methylidene complexes $\rm CH_2$-An-XY (X,Y = $\rm F_2$, FCl, $\rm Cl_2$) have been reported recently \cite{Li-Andrews-2007-AngewChem,Andrews-Li-2008-InorgChem}.
\begin{figure}[h] 
   \centering
   \includegraphics[scale=.2]{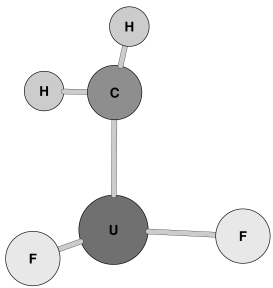} 
   \caption{Agostic interactions in a chiral uranium methylidene complex. Figure adapted from \cite{Andrews-Li-2008-InorgChem}}
   \label{fig:chiral-agostic}
\end{figure}
Initially searching for actinide complexes with An=C double bonds Andrews and coworkers found a class of compounds which displays strong agostic interactions and pyramidalization at the actinide atom leading to chiral structures. In agreement with the changes in the vibrational frequencies and NMR properties of the system which are also reported, the NBO analysis reveals a donor-acceptor interaction between the actinide atom and one of the two CH bonds. It is concluded that the low lying 6d and 5f orbitals form the foundation of the energetic preference of agostic geometries over the anagostic geometries.

Calculations on agostic interactions involving silicon, both as SiC-agostic interactions \cite{Clot-Mountford-2006-Organometallics} as well as SiH-agostic interactions \cite{Eisenstein-Lliedos-2006-Organometallics,Vyboishchikov-2006-ChemEurJ} have been performed recently.

Clot, Mountford and coworkers studied cationic imido titanium alkyls which are stabilized by a $\beta$-SiC agostic interaction \cite{Clot-Mountford-2006-Organometallics}. The NBO analysis of this system revealed that many contributions to donor-acceptor stabilizations are present. Specifically, $\rm Si_\beta$-$C_\gamma$ and the classical $\rm C_\gamma$-H contributions are present. However, the leading term in the second order pertubation treatment was the donation from a $\rm C_\alpha$-$Si_\beta$ bond to an empty 3d orbital of the Ti atom. Again, canting the Ti-$\rm CH_2$ bond away allows this interaction to take place.
This study also noted difficulties with the assignment of agostic interactions through the usually expected decrease of the $\rm ^1J_{CH}$ coupling constant. In this case intrinsic global changes in the electronic structure masked this effect, thus NMR measurements could be misleading in some cases.
In a later study it was reported that the $\rm AlMe_3$ and $\rm ZnMe_3$ adducts of the same sytem behave similarly but at the same time show CH agostic interactions for the binding methyl groups \cite{Clot-Mountford-2006-JACS}.

A novel system with a potentially linear agostic bond was recently reported \cite{Nielson-2004-Dalton}. Nielson and coworkers were able to crystallize a tantalum compound with a uniquely caged triphenylmethyl CH bond where the hydrogen atom points directly at the metal atom. In a detailed discussion of the results of an NBO analysis it is revealed that there is substantial donation of density from the CH bond to the metal consistent with the agostic mode of interaction. It is also noted that the interaction is substantially different from the common linear arrangement that hydrogen bonds take. While only a preliminary study, this work could provide fresh impulses by introducing a new class of agostic interaction. More work in this area is certainly warranted, the need for a topological analysis of the electron density was mentioned specifically. This could provide valuable insight into this complex situation.

There is a large number of reports in the recent literature that report agostic geometries in catalytic systems. Either in the ground state structure of the catalyst itself or in transition states during the reaction. These include arylpalladium halide complexes \cite{Buehl-Hartwig-2004-JACS}, molybdenum and tungsten catalysts used in olefin metathesis \cite{Coperet-Eisenstein-2006-Dalton}, salicyladiminato palladium and nickel complexes \cite{Liu-2007-Theochem} and zirconocene benzyl complexes for propylene polymerization \cite{Sassmannshausen-2007-EurJInorgChem}, hydrogen binding ruthenium complexes which are reported to bind $\rm sp^3$ and $\rm sp^2$ C-H agostic ligands \cite{vdBoom-Martin-Milstein-2004-InorgChimActa} and unsaturated hydride- and hydrocarbyl-molybdenum complexes \cite{Ruiz-2007-Organometallics}. In a recent study on CH bond activation by thorium and uranium complexes an interesting 5 centered agostic transition state was found \cite{Hay-Yang-2008-Organometallics}.

NBO theory has also been used to verify previously reported close CH$\cdots$metal approaches which had been labeled agostic \cite{Desiraju-2006-ChemComm}. In this case natural bonding orbitals could clearly show that the central copper atom in a copper ephedrine derivative does in fact have no empty orbital of the right symmetry for an agostic interaction to take place. Additionally, natural population analysis could show that the hydrogen atom in question actually becomes more positive compared to the un-coordinated species. This points to a hydrogen bonding type of interaction instead of an agostic bond. Later Desiraju and Thakur expanded on the difference between agostic interactions and hydrogen bonds using NBO as a platform for their calculations \cite{Desiraju-2007-Theochem}.

Although there is a clear preference for AIM theory for the treatment of agostic interactions among purely theoretically working groups \cite{Clot-Eisenstein-StructBond-2004} and some experimental groups \cite{Scherer-McGrady-2004-AngewChem,McGrady-2000-CoordChemRev}, NBO theory seems to be more common in combined x-ray and density functional studies, this is probably due to the relative ease of use and wide availability in standard quantum chemistry software.

\subsubsection{other methods}

Another commonly applied procedure to elucidate chemical bonding is the \emph{electron localization function} or \emph{ELF}. First introduced as a measure to compare the degree of electron localization in a given system compared to a uniform electron gas by Becke and Edgecombe \cite{BeckeEdgecombe-1990-JCP} and soon applied to chemical systems \cite{SavinBecke-1991-AngewChem,Savin-1992-AngewChem}. Soon after, Silvi and Savin introduced a topological analysis similar to AIM but based on the electron localization function instead of the electron density which can provide information about the shape, location and multiplicity of chemical bonds \cite{SilviSavin-1994-Nature,Silvi-2005-MhChem}

While this method suggests itself in the case of agostic interactions only three such publications have entered the literature to the best knowledge of the author.

Russo and coworkers report a study on the interaction of $\rm Mn^+$ in various electronic states with several small molecules in a dehydrogenation process \cite{Russo-2003-JPhysChemA}. The potential energy surface of the reaction path is calculated and ELF analyses of all intermediates and transition states are carried out. The evolution of the bonding along the reaction coordinate is followed describing the reaction mechanism in terms of agostic bond formation and breaking.

In a combined ELF and AIM study the fundamental differences between $\alpha$- and $\beta$-agostic interactions were highlighted \cite{Dobado-2006-Organometallics}. Both approaches yield similar results. To determine whether the agostic geometry is the result of a binding interaction analyses were performed on the minimum structure and as well on structures were an agostic geometry is forced on the molecule. The results show that in many cases the electronic effects that are reportedly the effect of an agostic interaction, such as a decrease in population of the CH bond and the atomic partial charge of the hydrogen atom, are in fact larger in artificially created agostic molecules. Thus, it is concluded, these effects cannot be the driving factor of agostic bond formation. The authors go as far as to suggest to drop the use of "agostic bond" in connection with $\alpha$-agostic interactions and to replace it with $\alpha$-agostic geometry to indicate that the short hydrogen-metal distance is a side effect of other interactions.

Alikhani and coworkers have investigated the insertion of early transition metals into the methane CH bond \cite{Alikhani-2008-ChemPhys}. The pathway of the insertion reaction is here reconstructed theoretically which ultimately lead to a dimethyl metal dihydride. The strength of the agostic bond in those complexes is found to decrease from Ti $>$ Zr $>$ Hf.

One of the oldest concepts in chemistry is to define the strength of a bond in terms of the force that is required to stretch the bond in question. While this seems intuitive at first it can be misleading since the force constants of the molecular vibrations depend on the curvature of the potential surface at the minimum while the dissociation energy of a bond or molecule depends on the relative energy of the extrema of the potential energy surface. Even in the case of isoelectronic species like $\rm N_2$ and CO this method is bound to fail since CO has a shallower potential but a higher dissociation energy. Thus, it seems unlikely that this method will yield even qualitative results. Nevertheless, this criterion was also applied in the investigation of several titanium and tungsten compounds showing agostic interactions \cite{Streubel-Grunenberg-2006-Organometallics}. The strength of the interaction was estimated to be smaller than 10 kcal/mol. This approach was also used in the computational investigation of a lanthanide compound but while a compliance constant was computed no estimate about the strength of the agostic interaction was made \cite{Tamm-2008-InorgChimActa}.

\subsection{Spectroscopic Approach}

\subsubsection{NMR properties}

One of the tell-tale signs of an agostic interaction is a significantly lower NMR $\rm ^1J_{CH}$ coupling constant. While the CH bond length is only marginally elongated, the coupling constant can decrease by more than 50\% \cite{Eisenstein-2006-Polyhedron}. Because of the computational effort that is usually involved in the calculation of NMR properties theoretical investigations for larger systems are only recently entering the literature \cite{Buehl-Hartwig-2004-JACS,Coperet-Eisenstein-2006-Dalton,Sassmannshausen-2007-EurJInorgChem,Clot-Mountford-2006-Organometallics,Vyboishchikov-2006-ChemEurJ}.

In a recent benchmark study the systematic calculation of NMR coupling constants was investigated \cite{Eisenstein-2006-Polyhedron}. Eisenstein and Solans-Montfort report that $\rm ^1J_{CH}$ coupling constants can be reproduced by relatively small basis-sets (6-31G(d,p)) on the respective carbon and hydrogen atoms, but overestimate the experimental value by about 20 Hz. However, the qualitative differences between the syn and anti isomers of the small rhenium, molybdenum and tantalum model compounds in question are retained. Including the actual ligands present in the experiment (tBu and $\rm CH_2tBu$ instead of $\rm CH_3$ and $\rm CH_2CH_3$) only decreases the coupling by 3-8 Hz depending on the system. This is consistent with earlier observations that the treatment of the local electron density, particularly of the electron shells close to the nucleus, is more important than an inclusion of the whole ligand environment as long as the bond lengths and angles in the model system are not changed significantly. Consequently when the representation of the valence orbitals was improved, little change for the CH coupling constants was observed after the inclusion of a diffuse function. However, when larger triple-$\zeta$ basis sets were employed (6-311++G(2df,2pd)), the value of the $\rm ^1J_{CH}$ constants decreased to values significantly lower than the experimental values. Much better agreement with experiment is obtained with the IGLO basis sets which have been developed specifically for the calculation of nuclear resonance data. In this case the calculated values agree with the experimental ones within less than 4 Hz for the small model. The improved performance of the IGLO basis sets is usually attributed to the better representation of the density close to the nuclei. However, experimental coupling constants of systems with strong agostic interactions seem to be less well reproduced by computation than those of systems with relatively weak agostic interactions.

These results were paralleled and, in essence, confirmed by others in an investigation of the intramolecular proton transfer of metalated ruthenium complexes. \cite{Gruendemann-2007-PNAS}. For this heavier system the CH coupling constant is calculated 2-10 Hz higher than the experimentally measured one. The strength of the agostic interaction was estimated from the activation barrier of an interconversion to be 24.9 kcal/mol.

In a more recent study a relation between $^{13}C$ chemical shift anisotropies and the strength of agostic interactions was investigated \cite{Emsley-2008-JACS}. This combined experimental solid state nmr and theoretical density functional study opens the interesting new area of using the orientation of the chemical shift tensor as a probe for agostic interactions.

A complete agreement of calculated NMR properties with the experiment cannot be expected because chemical shifts and coupling constants can be significantly dependent on the solvent and temperature -- effects usually not included in computational models.

\subsubsection{Vibrational Constants}

Because agostic bonds are usually slightly elongated compared to their anagostic counterparts the associated CH stretching frequency is lower than that of a CH bond with no additional interactions present.
This has been used in a series of combined experimental laser ablation and spectroscopic as well as computational studies of small carbene hydride systems $\rm H_2C$=$\rm MH_2$ (M = Ti, Zr, Hf, Mo, W) \cite{Andrews-2008-JPhysChemA,Andrews-Li-2008-InorgChem,Andrews-2008-InorgChimActa,Andrews-2006-ChemAsianJ,Andrews-2006-JPhysChemA,Andrews-2005-Organometallics,Andrews-2005-JACS,Andrews-2005-AngewChem,Andrews-2005-ChemEurJ,Zhou-2006-JACS,Andrews-2008-EurJInorChem,Liu-2008-JOrganometChem}. This combined approach does not only allow to probe systems for agostic interactions it also provides a valuable tool for the often difficult identification of compounds in matrix IR. By computational comparison of different isotopomers it was also possible to estimate the strength of the agostic interaction \cite{Andrews-2005-Organometallics} or to eliminate possible candidates for agostic interactions \cite{Andrews-2008-EurJInorgChem2,Andrews-2008-JACS}. In order to achieve this IRC calculations were used to follow the structure from one minimum to the other. Using this approach the interaction was estimated to be between 0.2 kcal/mol (B3LYP) and 2.3 kcal/mol (MP2) strong. Cop\'eret and Eisenstein show in a very interesting study of several olefin metathesis catalysts that a linear correlation between the stretching frequency of the agostic C-H bond and the $J_{\rm CH}$ NMR coupling constant exists
 \cite{Coperet-Eisenstein-2006-Dalton}. This is particularly useful for systems where one or the other experimental measurement might be difficult or impossible to carry out.

\section{Conclusions}

Given the wealth of publications in recent years it is obvious that agostic interactions still continue to draw attention and inspire further work. The very nature of the interaction is far from completely understood and many computational tools have been used to probe a wide variety of systems. It is the aim of this review to collate this information to make it accessible for others.

Still, some topics haven't been covered at length. QM/MM theory, which allows the treatment of much larger systems due to the separation of a smaller \emph{core} region which is treated quantum mechanically and a larger \emph{environment} which is treated with a molecular modeling approach, could be promising for the future. However, it has been reported that results obtained for agostic systems depend heavily on the exact method used \cite{Macgregor-2007-Organometallics} or that earlier results obtained with QM/MM could not be reproduced once the entire system was treated with high accuracy QM methods \cite{Maseras-Etienne-2006-Dalton,McGrady-2007-JCTC}.

More computational complications can arise when one wants to elucidate the strength of an agostic interaction. While several approaches have been mentioned already, the calculation through bond dissociation energies in bimolecular reactions can overestimate weak interactions such as agostic interactions if the BSSE is not taken into account \cite{Muckermann-Fujita-2007-JPhysChemB}.

Other areas like highly accurate multiconfigurational \emph{ab-initio} calculations \cite{Roos-Andrews-2007-JPhysChemA} or periodic boundary surface calculations \cite{Goddard-2004-JACS} also have to be mentioned only in passing.

It appears that even after may years of more and more experimental examples of agostic systems the jury is still out on what is the best way to probe a system theoretically for agostic interactions. The prevalent methods have been presented and discussed in this review to hopefully stimulate further computational enquiries into the true nature of this unique interaction.

\section{Acknowledgments}
The auhor wishes to thank Peter Schwerdtfeger (Massey) and Ralf Tonner (Massey) for comments and suggestions.


\end{document}